\documentstyle[aps,twocolumn]{revtex}
\begin{document} 
\title{Anderson-localization versus delocalization of interacting
       fermions in one dimension}
\author{P.\ Schmitteckert,${}^1$ T.\ Schulze,${}^2$ 
  C.\ Schuster,${}^2$ P.\ Schwab,${}^{2,3}$ and U.\ Eckern${}^2$\\[1.25ex]
\small
 { ${}^1$\em IPCMS-GEMME, CNRS, 23 rue du Loess, 
  F-67037 Strasbourg Cedex, France}\\
 { ${}^2$\em Institut f\"ur Physik, Universit\"at Augsburg, 
  D-86135 Augsburg, Germany}\\
 { ${}^3$\em Dipartimento di Fisica, Universit{\`a} 
       ``La Sapienza'', Piazzale A.\ Moro 2, I-00185 Roma, Italy}
}
\date{\today}
\maketitle
\begin{abstract}
Using the density matrix renormalization group algorithm,
we investigate the lattice model for spinless fermions in one
dimension in the presence of a strong
interaction and disorder. The phase sensitivity of the ground state
energy is determined with high accuracy for systems up to a size of
60 lattice constants. This quantity is found to be log-normally
distributed. The fluctuations grow algebraically with system size 
with a universal exponent of $\approx 2/3$ in the localized region of
the phase diagram. Surprisingly, we find, for an
attractive interaction, a delocalized phase of finite extension. The
boundary of this delocalized phase is determined.
\end{abstract}
\strut 
\vspace{1.0cm} \\
\noindent
The influence of electron-electron interaction on Anderson localization
has attracted a lot of interest for several years.
Many recent studies were 
motivated by the experimental observation of persistent currents in mesoscopic
rings
\cite{Levy90}. Motivated by an early suggestion \cite{Ambegaokar90} that the
interaction between the electrons may give a significant contribution to the
average persistent current, this phenomenon 
in the presence of both interaction and disorder 
has been investigated by various methods
\cite{Berkovits93,Bouzerar94,Giamarchi95,Ramin95,Kato94,Cohen96}. 
Nevertheless,
the magnitude of the effect is still not well understood.

In one dimension, interacting systems in the absence of disorder
\cite{Hamer87,Shastry90,Loss92}, 
as well as for disordered systems in the absence of interactions
\cite{Abrikosov78} are well studied. However, a clear understanding of 
the interplay between interaction and disorder has not yet been
obtained. In this Letter, we present novel results of a detailed study
of a simple interacting-fermion model with disorder. We determine the
ground state phase sensitivity with
high accuracy for a wide range of parameters and system sizes up to
60 lattice constants. Our main results are (i) a universal behavior of
the rms-value of the logarithmic phase sensitivity, which grows
with system size, $M$, proportional to $M^{2/3}$ in the localized region,
and (ii) the zero-temperature phase diagram, which shows, 
for an attractive interaction a delocalized phase of
finite extension. 

The numerical results are obtained with
the density matrix renormalization group algorithm
(DMRG) 
\cite{White92}, which allows calculation of 
ground state properties of disordered,
interacting fermion systems with an accuracy which is comparable to exact 
diagonalization, but for much larger systems \cite{Pang93,Peters96}. 
In our implementation of the DMRG we perform 
5 finite lattice sweeps keeping up to 750 states per block.

We consider a chain of spinless fermions with nearest-neighbor 
interaction and disorder,
\begin{eqnarray}\label{eq1}
 H &=&  -t \sum_{i} 
\left( c^+_ic^{\hphantom +}_{i+1} +c^+_{i+1}c^{\hphantom +}_{i} \right) 
       \;+\; \sum_i \epsilon^{}_i c^+_{i}c^{\hphantom +}_{i}   \nonumber \\
     && +\; V \sum_i n_i n_{i+1}
,\end{eqnarray}
and twisted boundary conditions, 
$c_0 = {\rm e}^{{\rm i} \phi }c_M$.
The length of the chain is denoted by $M$, and the particle number is
$N$. For simplicity, we will set $t=1$ in some of the formulas below.

The ground state energy $E(\phi)$ depends on the phase $\phi$.
The energy difference between periodic and anti-periodic boundary conditions,
$\Delta E= (-)^N \left[E(0) -E(\pi)\right]$, the persistent current,
$I(\phi) \sim -E'(\phi)$, and the charge stiffness,
$D \sim E''(\phi \! = \! 0)$, are a measure of the phase sensitivity of the
system.
In the clean limit, i.e.\ $\epsilon_n =0$ for all $n$,
the ground state energy can be determined from the Bethe Ansatz 
\cite{Hamer87,Eckle87}.
At half filling, the phase sensitivity in the limit of large 
systems ($M \to \infty$) is given by \cite{Remark}
\begin{equation}\label{eq6}
E_M (\phi)-M\varepsilon_\infty = - { \pi v \over 6M} 
\left( 1- 3K { \phi^2 \over \pi^2 } \right)
,\end{equation}
where $E_M$ is the ground state energy of the
$M$-site system, $\varepsilon_\infty$ is the energy 
density in the thermodynamic 
limit, and $v$ is the Fermi velocity, $v={\pi t\sin(2\eta)/( \pi -2\eta)}$.
The interaction parameter is $K=\pi / 4\eta$,
where $\eta$ parameterizes the interaction according to
$V=-2t\cos(2\eta)$.
Thus $M \Delta E = \pi v K /2$.

For the noninteracting system, the phase sensitivity in the presence of
a single defect ($\epsilon_0 \ne 0$) can be determined easily
\cite{Cheung88}.
In the presence of both electron-electron interaction and defect,
it is more difficult to calculate $\Delta E$.
However, it is known that the ground state and the low-lying excitations
can be described within the framework of a Luttinger liquid
\cite{Loss92,Gogolin94}.
Combining a first-order perturbative calculation with 
the scaling equations of Kane and Fisher \cite{Kane92},
we find in the case of a weak impurity
\begin{equation}\label{eq13}
 M \Delta E = {\pi v K \over 2} - |\epsilon_0 | 
\left( {M\over M_0} \right)^{1-K},
\end{equation} 
where $M_0$ is a short distance cut off which is -- for the half filled band -- 
of the order of the lattice spacing.
Using a duality relation between a weak 
impurity and a weak link \cite{Kane92},
we obtain for a strong impurity 
\begin{equation}\label{eq15}
M\Delta E  = {4t^2 \over |\epsilon_0| }\left( {M\over M_0}\right)^{1-1/K}
,\end{equation}
since the transmission through a strong defect is proportional to
$t^2/\epsilon_0$.
Figure \ref{fig1} shows $M\Delta E$ as a function of interaction
for system sizes $M=60$ and several defect strengths.
The points are numerical results from the DMRG, the full 
lines are analytical results
from the equations above. The cut-off parameter $M_0$ 
was fitted in order to obtain agreement
between numerical and analytical results, giving $M_0\approx 2$.     
An attractive interaction makes the barrier more transparent, 
while a repulsive interaction increases
the defect strength.
Deviations from the analytical results are found when 
the ``strong'' impurity becomes so weak
that a first order expansion is no longer appropriate. In addition, 
we find further
deviations near $V=\pm2$, where the Luttinger liquid becomes unstable.
At $V=-2$, there is an instability with respect to phase 
separation ($v \to 0, K \to \infty$).
At $V=+2$ there is an instability to formation of a 
charge-density wave, since at this point
$4k_F$-backscattering processes become relevant.

In the presence of a weak random potential, but $V=0$, we find,
generalizing the single impurity result,
\begin{equation}\label{eq16}
M\Delta E = \pi t - | \sum_{n=1}^M \epsilon_n {\rm e}^{2{\rm i} k_F n } |
,\end{equation}
where we introduce disorder by taking the $\{\epsilon_n\}$
uniformly random distributed over the interval 
$[-W/2, W/2]$.
At half filling, $k_F=\pi/2$, the sum
$
 \sum_{n=1}^M \epsilon_n (-1)^n
$
can be considered as a onedimensional random walk with $M$ steps. 
Recalling that a random walk leads to a Gaussian distribution
of distances, 
we obtain for the average phase sensitivity and the fluctuations
\begin{eqnarray}\label{eq20}
\langle M \Delta E \rangle &=& \pi  t- {   W \sqrt{M} \over \sqrt{6\pi} }, \\
\label{eq21}
\sigma_{M\Delta E}^2 &=& M {W^2 \over 12}   
\left( 1- {2\over \pi} \right)
.\end{eqnarray} 
The brackets $\langle \cdots \rangle $ denote the impurity average.
It is apparent in Eq.\ (\ref{eq20}) that perturbation 
theory breaks down for arbitrarily weak disorder, if the system is 
large enough: as is well known, in 1-D even weak
disorder leads to localization with a localization 
length which is proportional 
to $t^2/W^2$.
In large systems, $M\Delta E$ is drastically reduced due to 
disorder, however, numerical results (using exact diagonalization
methods, which is straightforward as long as $V=0$) indicate that
it remains positive for all realizations of the disorder 
\cite{Peters96}, 
in agreement with a 
theorem by Leggett \cite{Leggett91}.
For large systems we find an exponential
decay of the average phase sensitivity. 
In the localized regime, i.e.\ $M > \xi$,
the logarithm of $M\Delta E$ has approximately a normal distribution
\cite{Peters96}.
From our numerical data, where we averaged over 
$10^4$ realizations of the disorder potential, and
considered systems of up to $10^3$ sites, we find for
the average logarithmic phase sensitivity and its variance $\sigma^2$
in the limit of large systems ($M\to \infty$):
\begin{eqnarray}\label{eq23}
\langle \ln (M \Delta E )\rangle & \approx &  -{M\over \xi } + 0.76  \\
\label{eq24}
\sigma_{\ln (M\Delta E)} 
& \approx & 
\left( { 0.52 M \over\xi}\right)^{2/3}
,\end{eqnarray}
with $\xi =114t^2 /W^2$.  
In order to check the universality of the exponent,
we calculated the phase sensitivity for 
strong disorder up to $W=15t$, and for different fillings.
We always found the exponent $2/3$ in the localized region.

The interaction changes some of the results described above drastically.
Applying the Kane-Fisher scaling to Eq.\ (\ref{eq16}) we obtain
\begin{equation}\label{eq33}
M \Delta E = {\pi v K \over 2} - 
| \sum_{n=1}^M \epsilon_n {\rm e}^{2{\rm i} k_F n} | 
\left( {M \over M_0 } \right)^{1-K}
\end{equation}
since the strength of each defect is renormalized.
The average phase sensitivity is then given by
\begin{equation}\label{eq34}
\langle M \Delta E \rangle = {\pi v K \over 2} -  
{W \sqrt{M_0} \over \sqrt{6\pi} } 
\left( {M\over M_0} \right)^{(3-2K)/2}
,\end{equation}
and the fluctuations are
\begin{equation}\label{eq35}
\sigma_{M\Delta E}^2 
  = {W^2 M_0 \over 12}\left( 1-{2\over \pi} \right) 
\left( {M\over M_0} \right)^{3-2K}
.\end{equation}
Again, a repulsive interaction tends to
enhance the effective strength of the
defects, and an attractive interaction reduces it. 
Especially, for $K> 3/2$,
i.e.\ $V<-1$, the strength of each defect vanishes so fast 
that disorder becomes an irrelevant
perturbation: there is no localization \cite{Apel82,Giamarchi88}.  
We discuss the localized phase, $V>-1$, first.
Assuming that only one relevant length scale exists, 
i.e.\ the localization length $\xi$, 
one concludes from (\ref{eq34}) that   
$\xi \propto W^{2/(2K-3)}$ for weak
disorder. 
This is verified in Fig.\ \ref{fig2} 
where we plot the logarithmic phase 
sensitivity as a function of the scaled systems size. 
In the case of the largest systems considered ($M=60$), we averaged
over several hundred realizations, whereas for short systems ($M<20$)
we used ensembles of more than $10^3$ realizations. 
With good accuracy, points corresponding to different strengths 
of disorder lie on the same curve,
i.e.\ the localization length is indeed the only relevant 
scale, even for $M \gg \xi$, where perturbation theory
breaks down.
The average phase sensitivity, shown in Fig.\ \ref{fig2}, is for large
systems approximately given by ($V=1.2$)
\begin{equation}
\langle \ln (M \Delta E) \rangle =
- {M /\xi } + 1
,\end{equation}
with the localization length 
$\xi \approx 28 W^{-2/(3-2K)}$. 

The rms-value, $\sigma_{\ln (M\Delta E) }$, shown 
in Fig.\ \ref{fig3}, is for small systems
proportional to $M^{(3-2K)/2}$, see Eq.\ (\ref{eq35}). (Note that
$\sigma_{\ln (M\Delta E) }$ and $\sigma_{M\Delta E}$ are directly 
related to each other, provided 
$\sigma_{M\Delta E} \ll \langle M\Delta E \rangle $.)
For large systems we again find the fluctuations 
to be proportional to $M^{2/3}$, as 
in the noninteracting limit. Explicitly, we found from our numerical data
($V=1.2$, i.e.\ $K \approx 0.71$)
\begin{equation}
\sigma_{\ln (M \Delta E)} \approx 0.027 \left( M W^{2/(3-2K)} \right)^{2/3}
.\end{equation}

In Fig.\ \ref{fig4} we plot
$\langle \ln (M \Delta E )\rangle$
as a function of interaction and for several system sizes (here $W=1$). 
For comparison,
we included the phase sensitivity in the absence of disorder. Between
$V\approx -1.6$ and $\approx -1.1$,
the phase sensitivity remains almost
unreduced, even for large systems.
We believe that this region corresponds to the delocalized phase predicted
earlier \cite{Apel82,Giamarchi88}. This assertion is confirmed by an
apparent divergence of the localization length when approaching the
phase boundary from the localized side \cite{Schulze97}. Nevertheless,
the phase sensitivity remains smaller than 
in the clean system since the parameters $v$ and $K$ scale downwards 
due to the random potential 
\cite{Giamarchi88}.

The fluctuations of the logarithmic phase sensitivity provide another,
more accurate method for determining the extension of the delocalized
phase. Selected data are shown in Fig.\ \ref{fig5} for $M=10$ and 30.
As discussed in connection with Eq.\ (\ref{eq35}), a decreasing 
variance (with increasing $M$, compare $\Diamond$ with $+$)
implies that the disorder scales to smaller values, hence the 
system is delocalized, while the variance increases for a localized
ground state. Using this property as the criterion, we obtain, with
considerable numerical effort, the phase diagram shown in Fig.\
\ref{fig6}. These results are based on system sizes between 30 and
50 sites.
Clearly, by this method, we can only give a rough 
estimate of the phase boundary, and it is possible that we somewhat
overestimate the size of the delocalized region. 

In summary, using the DMRG algorithm, we have obtained high accuracy results 
for the ground state energy for a model
of interacting fermions with disorder. In the weak disorder limit,
we verified quantitatively several predictions 
on disordered Luttinger liquids. 
In the localized region, we determined
the localization length and the distribution of the phase sensitivity.
The latter is nearly log-normally distributed, with a universal 
size dependence of the fluctuations proportional to $M^{2/3}$.
We confirmed
the existence of a delocalized region in the phase diagram.
As far as we know, we are the first to give a quantitative 
estimate of the size (as a function of disorder and interaction)
of this region.

This work was supported by the Deutsche Forschungsgemeinschaft
(Forschergruppe HO 955/2-1) and through the TMR program of the European Union 
(P.\ Schm.\ and P.\ Schw.).
The calculations were partly done on the IBM
SP2 at the Leibniz-Rechenzentrum in Munich.
\newpage

\setlength{\unitlength}{0.1bp}
\begin{picture}(2519,1511)(0,0)
\put(1941,710){\makebox(0,0)[r]{$\epsilon_0=\hphantom{0.}6$}}
\put(1941,810){\makebox(0,0)[r]{$\epsilon_0=0.2$}}
\put(1941,910){\makebox(0,0)[r]{$\epsilon_0=\hphantom{0.}0$}}
\put(1468,-49){\makebox(0,0){interaction, $V$}}
\put(280,855){%
\makebox(0,0)[b]{\shortstack{phase sensitivity, $M\Delta E$}}%
}
\put(2257,151){\makebox(0,0){2}}
\put(2060,151){\makebox(0,0){1.5}}
\put(1863,151){\makebox(0,0){1}}
\put(1665,151){\makebox(0,0){0.5}}
\put(1468,151){\makebox(0,0){0}}
\put(1271,151){\makebox(0,0){-0.5}}
\put(1073,151){\makebox(0,0){-1}}
\put(876,151){\makebox(0,0){-1.5}}
\put(679,151){\makebox(0,0){-2}}
\put(540,1350){\makebox(0,0)[r]{3}}
\put(540,1167){\makebox(0,0)[r]{2.5}}
\put(540,984){\makebox(0,0)[r]{2}}
\put(540,801){\makebox(0,0)[r]{1.5}}
\put(540,617){\makebox(0,0)[r]{1}}
\put(540,434){\makebox(0,0)[r]{0.5}}
\put(540,251){\makebox(0,0)[r]{0}}
\end{picture}
\\[10ex]
FIG. \ref{fig1}: \refstepcounter{figure} \label{fig1}
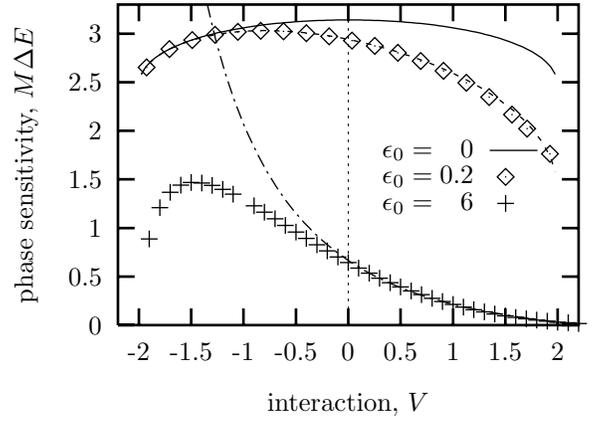
Phase sensitivity of the ground state energy 
in the presence of a single defect as a function
of interaction for several defect strengths $\epsilon_0$.
The diamonds and crosses are numerical results (system size $M=60$).
The straight lines are analytical results as described in the text.\\
\vspace{2truecm}
%
\setlength{\unitlength}{0.1bp}
\begin{picture}(2519,1511)(0,0)
\put(2011,1059){\makebox(0,0)[r]{$W=3$}}
\put(2011,1159){\makebox(0,0)[r]{$W=2$}}
\put(2011,1259){\makebox(0,0)[r]{$W=1$}}
\put(1468,-49){\makebox(0,0){scaled system size, $M \cdot W^{2/(3-2K)}$}}
\put(280,855){%
\makebox(0,0)[b]{\shortstack{$\langle \ln (M \Delta E) \rangle$}}%
}
\put(2336,151){\makebox(0,0){160}}
\put(2119,151){\makebox(0,0){140}}
\put(1902,151){\makebox(0,0){120}}
\put(1685,151){\makebox(0,0){100}}
\put(1468,151){\makebox(0,0){80}}
\put(1251,151){\makebox(0,0){60}}
\put(1034,151){\makebox(0,0){40}}
\put(817,151){\makebox(0,0){20}}
\put(600,151){\makebox(0,0){0}}
\put(540,1460){\makebox(0,0)[r]{1}}
\put(540,1259){\makebox(0,0)[r]{0}}
\put(540,1057){\makebox(0,0)[r]{-1}}
\put(540,856){\makebox(0,0)[r]{-2}}
\put(540,654){\makebox(0,0)[r]{-3}}
\put(540,453){\makebox(0,0)[r]{-4}}
\put(540,251){\makebox(0,0)[r]{-5}}
\end{picture}
\\[10ex]
FIG. \ref{fig2}: \refstepcounter{figure} \label{fig2}
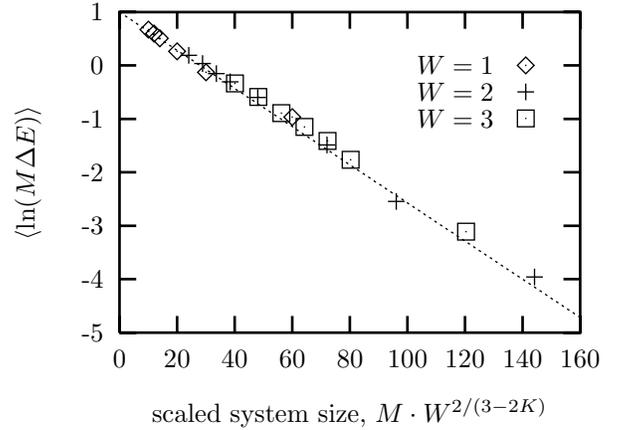
Average logarithmic phase sensitivity as a function
of the scaled system size, for $V=1.2$ and disorder $W=1,2,3$.\\
\newpage
%
\setlength{\unitlength}{0.1bp}
\begin{picture}(2519,1511)(0,0)
\put(2080,514){\makebox(0,0)[r]{$W=3.0$}}
\put(2080,614){\makebox(0,0)[r]{$W=1.0$}}
\put(2080,714){\makebox(0,0)[r]{$W=0.6$}}
\put(2080,814){\makebox(0,0)[r]{$W=0.4$}}
\put(2080,914){\makebox(0,0)[r]{$W=0.2$}}
\put(1468,-49){\makebox(0,0){scaled system size, $M \cdot W^{2/(3-2K)}$}}
\put(280,855){%
\makebox(0,0)[b]{\shortstack{fluctuations, $\sigma_{ \ln (M \Delta E )}$}}%
}
\put(1886,151){\makebox(0,0){100}}
\put(1243,151){\makebox(0,0){10}}
\put(600,151){\makebox(0,0){1}}
\put(540,1096){\makebox(0,0)[r]{1}}
\put(540,492){\makebox(0,0)[r]{0.1}}
\end{picture}
\\[10ex]
FIG. \ref{fig3}: \refstepcounter{figure} \label{fig3}
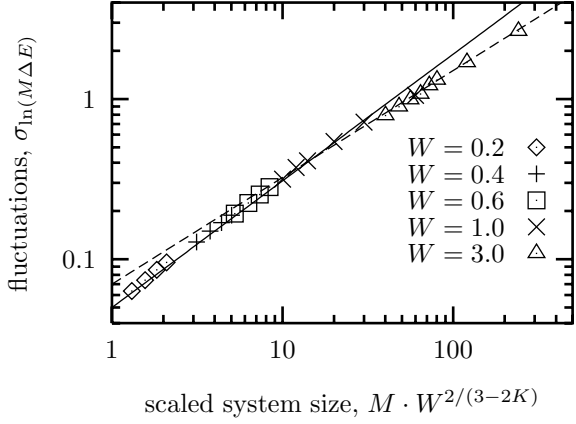
Rms-value of $\ln (M \Delta E)$ as a function of scaled system
size (again $V=1.2$). The full line is the analytic result 
according to Eq.\ (\ref{eq35}),
which explains the low-$M$ behavior.  
For large systems, 
$\sigma_{\ln (M\Delta E)}$ is proportional to $M^{2/3}$ (dashed line)
as in the noninteracting case.\\
\vspace{2truecm}
%
\setlength{\unitlength}{0.1bp}
\begin{picture}(2519,1511)(0,0)
\put(1245,455){\makebox(0,0)[r]{$M=40$}}
\put(1245,555){\makebox(0,0)[r]{$M=30$}}
\put(1245,655){\makebox(0,0)[r]{$M=20$}}
\put(1245,755){\makebox(0,0)[r]{$M=10$}}
\put(1468,-49){\makebox(0,0){interaction, $V$}}
\put(280,855){%
\makebox(0,0)[b]{\shortstack{$\langle \ln (M\Delta E) \rangle$}}%
}
\put(2336,151){\makebox(0,0){1.5}}
\put(2088,151){\makebox(0,0){1}}
\put(1840,151){\makebox(0,0){0.5}}
\put(1592,151){\makebox(0,0){0}}
\put(1344,151){\makebox(0,0){-0.5}}
\put(1096,151){\makebox(0,0){-1}}
\put(848,151){\makebox(0,0){-1.5}}
\put(600,151){\makebox(0,0){-2}}
\put(540,1460){\makebox(0,0)[r]{1.2}}
\put(540,1259){\makebox(0,0)[r]{1}}
\put(540,1057){\makebox(0,0)[r]{0.8}}
\put(540,856){\makebox(0,0)[r]{0.6}}
\put(540,654){\makebox(0,0)[r]{0.4}}
\put(540,453){\makebox(0,0)[r]{0.2}}
\put(540,251){\makebox(0,0)[r]{0}}
\end{picture}
\\[10ex]
FIG. \ref{fig4}: \refstepcounter{figure} \label{fig4}
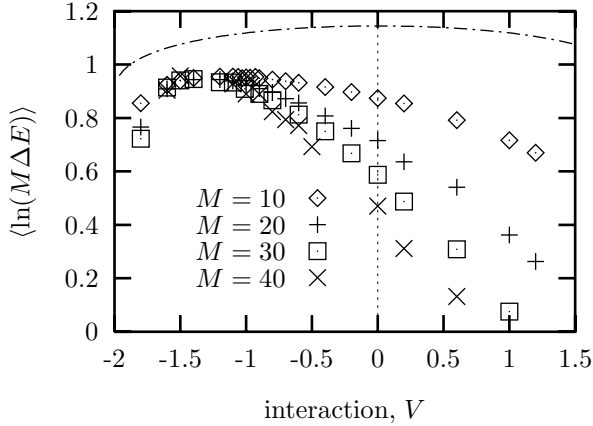
Average logarithmic phase sensitivity as a function of interaction, 
for system sizes ranging from 10 to 40; $W=1$.
For comparison, we included the result in the clean limit
($W=0$, dashed-dotted line).\\
\newpage
%
\setlength{\unitlength}{0.1bp}
\begin{picture}(2519,1511)(0,0)
\put(1381,1021){\makebox(0,0)[r]{$M=30$}}
\put(1381,1121){\makebox(0,0)[r]{$M=10$}}
\put(1468,-49){\makebox(0,0){interaction, $V$}}
\put(280,855){%
\makebox(0,0)[b]{\shortstack{fluctuations, $\sigma_{\ln (M \Delta E)}$}}%
}
\put(2162,151){\makebox(0,0){-0.2}}
\put(1815,151){\makebox(0,0){-0.6}}
\put(1468,151){\makebox(0,0){-1}}
\put(1121,151){\makebox(0,0){-1.4}}
\put(774,151){\makebox(0,0){-1.8}}
\put(540,1460){\makebox(0,0)[r]{0.25}}
\put(540,1218){\makebox(0,0)[r]{0.2}}
\put(540,976){\makebox(0,0)[r]{0.15}}
\put(540,735){\makebox(0,0)[r]{0.1}}
\put(540,493){\makebox(0,0)[r]{0.05}}
\put(540,251){\makebox(0,0)[r]{0}}
\end{picture}
\\[10ex]
FIG. \ref{fig5}: \refstepcounter{figure} \label{fig5}
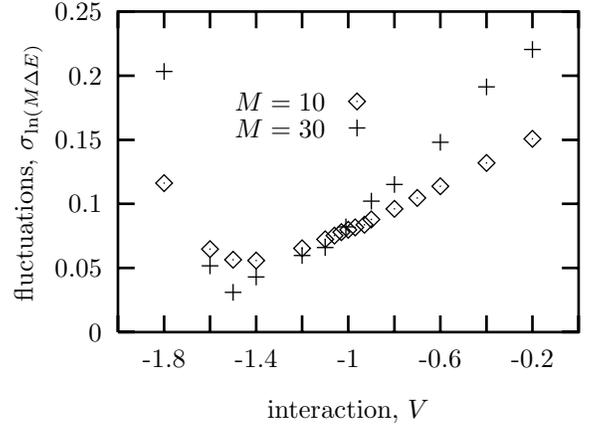
Rms-value of the logarithmic phase sensitivity 
versus interaction, for $M=10$ and 30; $W=1$.
\vspace{2truecm}
%
\setlength{\unitlength}{0.1bp}
\begin{picture}(2519,1511)(0,0)
\put(1468,-49){\makebox(0,0){interaction, $V$}}
\put(280,855){%
\makebox(0,0)[b]{\shortstack{disorder, $W$}}%
}
\put(2215,151){\makebox(0,0){0}}
\put(1811,151){\makebox(0,0){-0.5}}
\put(1407,151){\makebox(0,0){-1}}
\put(1004,151){\makebox(0,0){-1.5}}
\put(600,151){\makebox(0,0){-2}}
\put(540,1396){\makebox(0,0)[r]{1.8}}
\put(540,1269){\makebox(0,0)[r]{1.6}}
\put(540,1142){\makebox(0,0)[r]{1.4}}
\put(540,1015){\makebox(0,0)[r]{1.2}}
\put(540,887){\makebox(0,0)[r]{1}}
\put(540,760){\makebox(0,0)[r]{0.8}}
\put(540,633){\makebox(0,0)[r]{0.6}}
\put(540,506){\makebox(0,0)[r]{0.4}}
\put(540,378){\makebox(0,0)[r]{0.2}}
\put(540,251){\makebox(0,0)[r]{0}}
\end{picture}
\\[10ex]
FIG. \ref{fig6}: \refstepcounter{figure} \label{fig6}
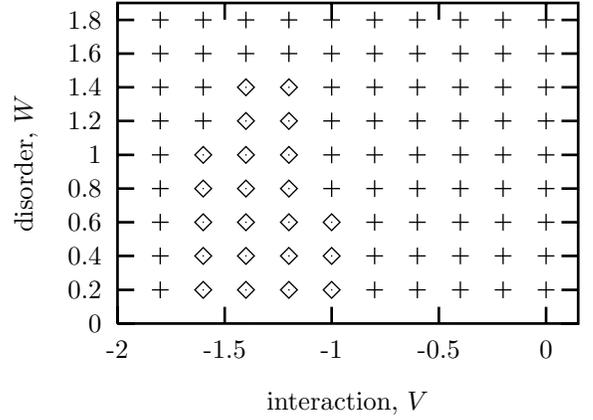
Phase diagram. The symbol $\Diamond$ ($+$) denotes the 
region where the variance of the logarithmic
phase sensitivity decreases (increases) as a function of the system
size. We considered up to $50$ sites. The $\Diamond$-region corresponds to
a delocalized ground state.


\begin{thebibliography}{10}
\bibitem{Remark}
The ground state energy is periodic in $\phi$ with period $2\pi$. Here we 
consider odd $N$, and the interval $-\pi\dots\pi$.
\end{thebibliography}
\end{document}